\documentstyle[prb,twocolumn,aps]{revtex}
\include{psfig}

\begin{document}
\title{\bf Proximity Effect and Multiple Andreev Reflections in
  Chaotic Josephson junctions}
\author{P. Samuelsson, G. Johansson, \AA. Ingerman, V.S. Shumeiko, and
  G.  Wendin} 
\address{Department of Microelectronics and Nanoscience,
  Chalmers University of Technology and G\"{o}teborg University,
  \\S-41296 G\"{o}teborg, Sweden} \date{\today} \maketitle
\begin{abstract}
  We study the dc-current transport in a voltage biased
  superconductor-chaotic dot-superconductor junction with an induced
  proximity effect(PE) in the dot. It is found that for a Thouless
  energy $E_{Th}$ of the dot smaller than the superconducting energy gap
  $\Delta$, the PE is manifested as peaks in the differential
  conductance at voltages of order $E_{Th}$ away from the even
  subharmonic gap structures $eV \approx 2(\Delta\pm E_{Th})/2n$.
  These peaks are insensitive to temperatures $kT \ll \Delta$ but are
  suppressed by a weak magnetic field. The current for suppressed PE
  is independent of $E_{Th}$ and magnetic field and is shown to be
  given by the Octavio-Tinkham-Blonder-Klapwijk theory.
  \cite{Octavio83}
\end{abstract}
Over the last decades, there has been a large interest in various
aspects of the proximity effect (PE) in mesoscopic normal
conductor-superconductor (NS) systems. The PE can on a microscopic
level be viewed as correlations on the scale of the Thouless energy
$E_{Th}$ between electrons and holes in the normal conductor, induced
via Andreev reflections at the NS-interface. Well known manifestations
of the PE are the bias conductance anomaly \cite{Kastalsky91} in
diffusive NS-junctions and the induced gap in the spectrum of
diffusive\cite{Golubov89} or chaotic\cite{Melsen96} SNS- and
NS-junctions.

There is however no complete theory of the PE in voltage biased
SNS-junctions. The reason is that electrons and holes in the normal
conductor undergo multiple Andreev
reflections\cite{Octavio83,Klapwijk82} (MAR), giving rise to
complicated correlation effects and strong nonequilibrium in the
normal part of the junction. \cite{Bezuglyi99} The fingerprint of MAR
transport is subharmonic gap structures (SGS) in the current-voltage
characteristics at $eV=2\Delta/n$.

So far only some limiting cases have been considered. In short
junctions, where $E_{Th}$ is much larger than the superconducting
energy gap $\Delta$, a coherent MAR-theory \cite{Bratus95} has been
developed which fully incorporates the PE.  The theory describes to
large accuracy experiments in atomic size point contacts
\cite{Vanderpost94} and has also been applied to short diffusive
junctions \cite{Bardas97} and disordered tunnel barriers.
\cite{Naveh00} For junctions in the short limit, the PE modifies the
shape but not the position of the SGS, as well as gives rise to
ac-Josephson effect. Other cases studied are long, $E_{Th} \ll
\Delta$, diffusive SNS-junction with no proximity coupling between the
superconductors \cite{Bezuglyi00} and SNS-junctions consisting of two
weakly coupled proximity NS-junctions in equilibrium.\cite{Aminov96}

Taken together, present theories do not give an answer to the general
question of the role of the PE in SNS-junctions where the Thouless
energy is comparable to the superconducting energy gap, $E_{Th}\sim
\Delta$. Moreover, recent experiments on SNS-junctions in this
regime\cite{Kutchinsky97} show qualitatively new features: splitting
of the SGS conductance peaks at even subharmonics $eV=\Delta$ and
$\Delta/2$.

In this Letter we study current transport in a superconductor-chaotic
dot-superconductor (S-dot-S) junction. By employing the scattering
theory of MAR\cite{Beenakker91} and using random matrix
theory\cite{Beenakker97} to describe the statistical properties of the
energy- and magnetic field dependent scattering matrix of the dot, we
are able to calculate the dc-current for arbitrary ratio between
$E_{Th}$ and $\Delta$ as well as for arbitrary strength of the PE due
to variation of the magnetic field in the dot.\cite{Melsen96}

The main result of the paper is the explanation of the splitting of
the even SGS conductance peaks, i.e the additional peaks at $eV
\approx 2(\Delta\pm E_{Th})/2n$, in terms of an induced PE in the dot.
The PE conductance peaks are suppressed by a weak magnetic field but
are insensitive to temperatures $kT \ll \Delta$.
\begin{figure}[h]
\centerline{\psfig{figure=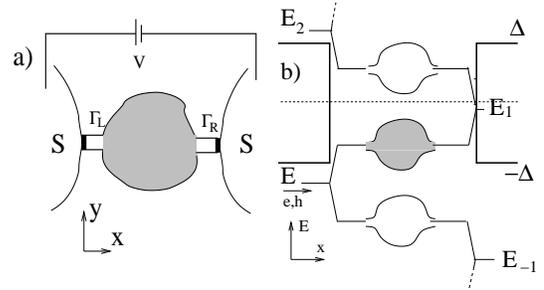,width=7.0cm}} 
\caption{a) Schematic picture of
  the junction. The shaded region is the chaotic dot and the black
  bars denote tunnel barriers with transparency $\Gamma_L$ and
  $\Gamma_R$.  b) A MAR-trajectory in energy-position space for
  quasiparticles injected at energy $E$ from left. Filled (empty)
  chaotic dots denote electrons (holes) scattering. Scattering
  processes in which quasiparticles are Andreev reflected at energies
  $E_{Th}$ away from the Fermi energy have an enhanced amplitude,
  giving rise to additional peaks in the conductance at $eV\approx
  2(\Delta\pm E_{Th})/2n$.}
\label{fig1}
\end{figure}
The junction is shown schematically in Fig.\ref{fig1}: A
two-dimensional quantum dot is coupled to two superconducting
electrodes via quantum point contacts supporting $N$ transverse modes
each. The contacts contain tunnel barriers with mode-independent
transmission probabilities $\Gamma_L,\Gamma_R \gg 1/N$. It is assumed
that the quasiparticle dwell time in the dot, $\hbar/E_{Th}$, is much
smaller than the inelastic scattering time; here
$E_{Th}=N(\Gamma_L+\Gamma_R)\delta/(2\pi)$ and $\delta$ is mean level
spacing of the dot. In this case the transport through the junction
can be characterized by the Andreev reflection amplitudes at the
normal lead-insulator-superconductor (NIS) interfaces and the
scattering matrix of the dot. We consider the case where the classical
motion in the dot is chaotic on time scales longer than the ergodic
time $\tau_{erg} \sim L/v_F$ ($l\gg L$) or $L^2/(lv_F)$ ($l \ll L$),
where $L$ and $l$ are the linear dimension and the mean free path of
the dot. The ergodic time is assumed to be smaller than the
quasiparticle dwell time and the inverse superconducting gap,
$\tau_{erg}\ll \hbar/E_{Th},\hbar/\Delta$. In this case random matrix
theory\cite{Beenakker97} can be used to describe the scattering
properties of the dot.

The scattering matrix $S$ can be written in terms of the Hamiltonian
$H$ (dimension $M$) of the closed dot as
\begin{equation}
S=\left( \begin{array}{cc} r & t' \\ t & r' \end{array} \right)=1-2\pi i W^{\dagger}(E-H+i\pi WW^{\dagger})^{-1}W,
\label{hamscat}
\end{equation}
where $r,t,r'$ and $t'$ are the $N \times N$ reflection and
transmission matrices and $W$ describes the coupling of the dot to the
leads, with $W_{nm}=\delta_{nm}(M\delta)^{1/2}/\pi$. The Hamiltonian
$H$ is described by a random Hermitian matrix $H=H_0+i\gamma H_1$,
where $H_0$ and $H_1$ are real symmetric and anti-symmetric matrices
respectively, independently distributed with the same Gaussian
distribution, $P(H_{0(1)})\propto
\mbox{exp}[-\pi^2(1+\gamma^2)\mbox{tr}(H_{0(1)}H_{0(1)}^T)/(4M\delta^2)]$.
The parameter $\gamma$ is related to the magnetic flux in the dot as
$\Phi \simeq \gamma \Phi_0(M\delta \tau_{erg}/\hbar)^{1/2}$,where
$\Phi_0=h/e$ is the flux quantum. A magnetic flux $\Phi_c\simeq
\Phi_0(\tau_{erg}E_{Th}/\hbar)^{1/2}$ effectively breaks time reversal
symmetry in the dot.

The current is calculated within a scattering approach for the
Bogoliubov-de Gennes equations \cite{Beenakker91} and is expressed in
terms of the scattering state wavefunctions $\Psi_{\sigma}$, as
$I=(e/h)\int dy
\sum_{\sigma}\mbox{Im}(\Psi_{\sigma}^{\dagger}[d\Psi_{\sigma}/dx])f_0$,
where $\sigma=\{e/h,E,L/R,j\}$ labels the scattering state
(electron/hole like quasiparticle injected at energy $E$ from the
left/right superconducting reservoir, in transverse mode $j$) and
$f_0$ is the equilibrium Fermi distribution of the reservoirs. The
energies are measured relative to the Fermi energy in the superconductors
and the magnetic field in the contacts is assumed to be zero.

The acceleration of the injected quasiparticles due to the applied
voltage $V$ gives a scattering state wavefunction which is a
superposition of electron and hole states at different
energies\cite{Beenakker91} $E_n=E+2neV$. For quasiparticles incident
from the left superconductor, the wavefunction on the normal side of
the left NIS-interface has the form
\begin{eqnarray}
\Psi_L&=& \sum_{m=1}^N\Phi_m(y)\sum_n e^{-iE_{2n}t/\hbar} \nonumber \\
&\times&\left[\begin{array}{c}(c_{m,2n}^{e,+}e^{ik_{m,2n}^ex}+c_{m,2n}^{e,-}e^{-ik_{m,2n}^ex})/(k_{m,2n}^e)^{1/2}\\ (c_{m,2n}^{h,+}e^{ik_{m,2n}^hx}+c_{m,2n}^{h,-}e^{-ik_{m,2n}^hx})/(k_{m,2n}^h)^{1/2} \end{array} \right] 
\label{wfansatz}
\end{eqnarray}
where $m$ is transverse mode index. At the right interface the
electron/hole energies are shifted by $\pm eV$ and the wave function
$\Psi_R$ is given by multiplying $\Psi_L$ by $\mbox{exp}(-i\sigma_z
eVt/\hbar)$ and substituting $2n \rightarrow 2n+1$ ($\sigma_z$ is the
Pauli matrix in electron-hole space). The wave vector
$k_{m,n}^{e(h)}=[(2m/\hbar^2)(E_F-\epsilon_m\pm E_n)]^{1/2}$, where
$\epsilon_m$ is the transverse mode energy and $+/-$ corresponds to
electrons/holes. The vector potential enters only the transverse
wavefunctions $\Phi_m$, normalized for each mode to carry the same
current. The scattering at the dot connects the electron and hole
wavefunction coefficients as
\begin{eqnarray}
\left(\begin{array}{c} {\hat c}_{n}^{e,-} \\ {\hat c}_{n+1}^{e,+} \end{array} \right)=S^+_n \left(\begin{array}{c} {\hat c}_{n}^{e,+} \\ {\hat c}_{n+1}^{e,-}\end{array} \right),
\left(\begin{array}{c}{\hat c}_{n}^{h,+} \\ {\hat c}_{n-1}^{h,-} \end{array} \right)=S^-_n \left(\begin{array}{c} {\hat c}_{n}^{h,-} \\ {\hat c}_{n-1}^{h,+}\end{array} \right),
\label{scatmat}
\end{eqnarray}
where we have introduced the vector notation ${\hat
  c}_n^{e,+}=[c_{1,n}^{e,+},...,c_{N,n}^{e,+}]$ and $S_{n}^+=S(E_{n})$
and $S_{n}^-=S^*(-E_{n})$. At the left NIS-interface, the scattering
is described by Andreev and normal reflection and transmission
amplitudes for electrons (e) and holes (h), $a^{e/h}_n=a^{e/h}(E_n)$ and
similarly $b^{e/h}_n,c^{e/h}_n$ and $d^{e/h}_n$, given in Ref.
\onlinecite{Blonder82}. This gives the connection between wavefunction
coefficients as
\begin{eqnarray}
\left(\begin{array}{c} {\hat c}_{n}^{e,-} \\ {\hat c}_{n}^{h,+} \end{array} \right)&=&\left(\begin{array}{cc} b^e_n & a^h_n\\ a_n^e & b^h_n \end{array} \right) \left(\begin{array}{c} {\hat c}_{n}^{e,+} \\ {\hat c}_{n}^{h,-} \end{array} \right)+\delta_{n0}\delta_{mj}\left(\begin{array}{c} c^e_n\\ d^e_n \end{array} \right),
\label{arcoff}
\end{eqnarray} 
where the source term ($\propto \delta_{n0}\delta_{mj}$) describes
electron quasiparticle injection. The coefficients at the right
NIS-interface are connected in a similar way. The other scattering
states are constructed analogously.

\begin{figure}[h]
\centerline{\psfig{figure=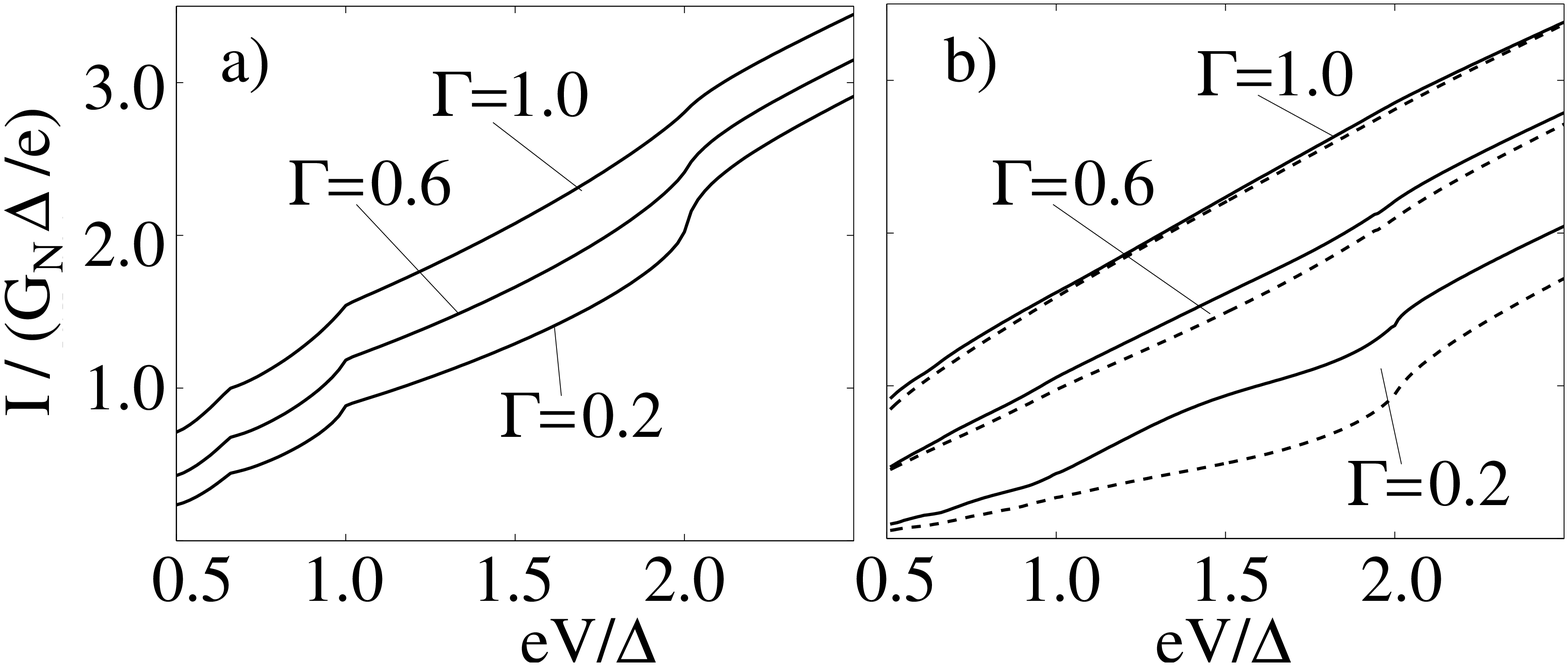,width=9cm}}
\caption{The current-voltage characteristics for a symmetric junction $\Gamma=\Gamma_L=\Gamma_R$ with a temperature $kT\ll\Delta$. a) In the short dwell time regime, $E_{Th} \gg \Delta$, the current show SGS at $eV=2\Delta/n$. b) In the intermediate dwell time regime, $E_{Th}=0.4\Delta$, the current is shown for $\Phi=0$ (solid) and $\Phi \gg \Phi_c$ (dashed) and $N=10$. The SGS is less pronounced and the current for low transparencies is smaller compared to the short dwell time regime.}
\label{fig2}
\end{figure}

In the short dwell time regime, $E_{Th} \gg \Delta$, the scattering
matrix is independent of energy on the scale of $\Delta$, and the
current can be written\cite{Bardas97,Naveh00} as a sum of the single
mode currents \cite{Bratus95}, with different transmission eigenvalues
$D_m$ (eigenvalues of the matrix product $~tt^{\dagger}$). The
ensemble averaged current $\langle I \rangle$ is then found via an
integration over transmission eigenvalues with the distribution
\cite{Brouwer96}
$\rho(D)=N/\pi[\Gamma(2-\Gamma)]/([\Gamma^2+4D(1-\Gamma)]\sqrt{D(1-D)})$.
The current voltage characteristics in Fig. \ref{fig2}a show SGS
at $eV=2\Delta/n$ and an excess current \cite{excesscurr} for all
$\Gamma$ [the normal state conductance
$G_N=(2e^2/h)N\Gamma_L\Gamma_R/(\Gamma_L+\Gamma_R)$].
For a longer quasiparticle dwell time, $E_{Th} \alt \Delta$, the
ensemble averaged current is calculated numerically by generating a
large number of Hamiltonians.\cite{numcom} The current voltage
characteristics is shown in Fig. \ref{fig2}b for $E_{Th}=0.4\Delta$.
Compared to the short dwell time regime in Fig \ref{fig2}a, the SGS is
less pronounced and the current for low barrier transparencies is
reduced for $\Phi=0$ and even further for $\Phi \gg \Phi_c$.

\begin{figure}[h]
\centerline{\psfig{figure=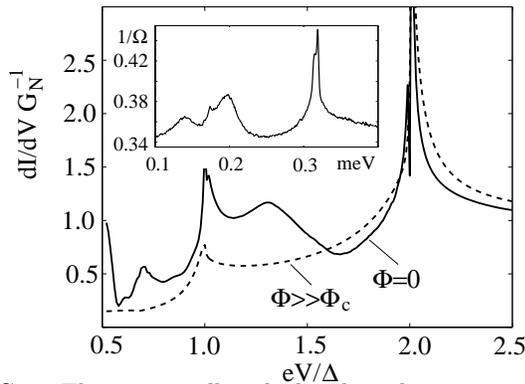,width=7.0cm}}
\caption{The numerically calculated conductance as a function of voltage for $\Gamma_L=0.1, \Gamma_R=0.3$, $E_{Th}=0.4\Delta$, $kT \ll \Delta$ and $N=15$. Inset: The differential conductance as a function of voltage for the diffusive SNS-junction, $L=0.29\mu m$ and $\Delta=0.17 meV$ at $kT=240 mK$, studied experimentally by Kutchinsky et al.\cite{Kutchinsky97} The theoretical and experimental curves show the same qualitative features with conductance peaks at $eV=2\Delta,\Delta$ and $\Delta \pm E_{Th}$.}
\label{fig3}
\end{figure}

The current voltage characteristics can be studied in detail by
considering the conductance, $dI/dV$, shown in Fig. \ref{fig3}
(parameters chosen to clearly display PE-features). The conductance
with suppressed PE, $\Phi \gg \Phi_c$, shows SGS at $eV=2\Delta/n$.
The PE is manifested as an enhancement of the SGS at $eV=\Delta$ but
also as additional peaks\cite{peaks} in the conductance on both sides
of $eV=\Delta$. It is found by varying $E_{Th}$ that the positions of
the additional peaks are given by $eV \approx \Delta \pm E_{Th}$.
Moreover, the position and magnitude of the peaks are found to be
insensitive to temperatures $kT \ll \Delta$.
These features of the conductance have recently been
observed\cite{Kutchinsky97,other} in experiments with diffusive
SNS-junctions of intermediate length, $E_{Th} \alt \Delta$
(shown as inset in Fig. \ref{fig3}). The experimental
curves show conductance peaks at $eV \approx 2\Delta,\Delta,\Delta \pm
E_{Th}$ and $eV \approx (\Delta \pm E_{Th})/2$, with position and
amplitude insensitive to temperatures $kT \ll \Delta$. In a
corresponding interferometer setup\cite{Kutchinsky97} the peaks at $eV
\approx \Delta \pm E_{Th}$ are demonstrated to result from induced PE
(the peaks at $eV \approx (\Delta \pm E_{Th})/2$ could not be
resolved).  The amplitude of all SGS peaks is successively reduced for
increasing junction length, and the peaks at $eV \approx \Delta \pm
E_{Th}$ are washed out in long junctions, $E_{Th} \ll \Delta$.
Although the system studied experimentally is different from a S-dot-S
junction, the two type of junctions can be expected to show
qualitatively similar behavior, in the same way as for an NS and a
N-dot-S junction (see Ref. \onlinecite{Clerk00}).

A qualitative understanding of the origin of the PE conductance peaks
can be obtained by first considering the peak at $eV
\approx\Delta+E_{Th}$. It is built up by quasiparticles which are
Andreev reflected once at energies of order of $E_{Th}$ away from the
Fermi energy (see Fig. \ref{fig1}b). These scattering processes have
enhanced amplitude due to the PE and are similar to the ones giving
rise to the finite bias conductance anomaly in
NS-junctions.\cite{Kastalsky91} The conductance peak is insensitive to
temperatures $kT \ll \Delta$, since the distribution of injected
quasiparticles from the superconductors is unaffected by temperatures
well below the superconducting gap, in contrast to the NS-conductance
anomaly, which is suppressed for $kT \gg E_{Th}$.

The conductance peak at $eV \approx \Delta-E_{Th}$ results from
scattering processes which include two Andreev reflections, with one
of the reflections at an energy of order $E_{Th}$ away from the Fermi
energy. These process, in the same way, have an enhanced amplitudes
due to the PE. Since the processes with one and two Andreev
reflections are different, the shape and amplitude of the conductance
peaks at $eV\approx \Delta \pm E_{Th}$ are in general different.
Following the same line of reasoning, we predict that the conductance
(for $E_{Th}$ smaller than the distance between subsequent subgap
harmonics) will show peaks at all $eV \approx 2(\Delta \pm
E_{Th})/(2n)$.

It follows from the quantum mechanical current expression that the
ensemble averaged dc-current can be written as $\langle I
\rangle=N(e/h)\sum_{\sigma,n}(f_{n,\sigma}^{e,+}-f_{n,\sigma}^{e,-}+f_{n,\sigma}^{h,+}-f_{n,\sigma}^{h,-})$,
where $f_{n}^{e/h,\pm}=\langle \mbox{tr}[({\hat
  c}_{n}^{e/h,\pm})^{\dagger}{\hat c}_{n}^{e/h,\pm}] \rangle f_0/N$
(suppressing index $\sigma$) are correlation functions of the
wavefunction coefficients and the trace is taken over the transverse
modes. The functions $f_{n}^{e/h,\pm}$ can be interpreted as ensemble
averaged distribution functions at energy $E_n$ for electron/hole
quasiparticles with positive/negative sign of the wavenumber, summed
over transverse modes. For broken time reversal symmetry in the dot,
$\Phi \gg \Phi_c$, it is possible to formulate matching equations for
the distribution functions $f_{n}^{e/h,\pm}$ directly, in the
following way:

We first note that for quasiparticles propagating in energy-position
space along the MAR-ladder (see Fig. \ref{fig1}b), the scattering
matrices of the dot at different energies are effectively
uncorrelated to leading order in $N\Gamma_L,N\Gamma_R$, (i.e.
neglecting quantum corrections). This result is an extension of what
is found for an N-dot-S junction, \cite{Brouwer96} by employing the
same diagrammatic technique for integration over the unitary group.

The statistical independence of the scattering matrices leads to three
rules for wavefunction coefficient correlations and averages: i)
coefficients with different energy indices $n$ are uncorrelated since
they are connected via at least one traversal through the dot. (This
also means that there is no ac-Josephson current.) ii) coefficients for
quasiparticles incoming towards the NIS-interface are uncorrelated,
because incoming quasiparticles have scattered at dots at different
energies before approaching the NIS-interface. iii) the average of any
coefficient itself is zero. With these rules for the correlations
between different wavefunction coefficients, we can, directly from the
matching Eqs. (\ref{scatmat}) and (\ref{arcoff}), derive matching
equations for the functions $f_{n}^{e/h,\pm}$. For the scattering
across the dot, e.g. for left injected electrons, we get from Eq.
(\ref{scatmat})
\begin{eqnarray}
f_{n}^{e,-}&=&\langle \mbox{tr}\left[(r {\hat c}_{n}^{e,+}+t'{\hat c}_{n+1}^{e,-}) \times c.c. \right] \rangle=1/2[f_{n}^{e,+}+ f_{n+1}^{e,-}] \nonumber \\
f_{n+1}^{e,+}&=&1/2 [f_{n+1}^{e,-}+f_{n}^{e,+}].
\label{match1}
\end{eqnarray}
In this derivation we used that the averaging rules gives $\langle
\mbox{tr}[({\hat c}_{n+1}^{e,-})^{\dagger}t'^{\dagger}t'{\hat
  c}_{n+1}^{e,-}]\rangle=\langle \mbox{tr}(t'^{\dagger}t')\rangle
\langle \mbox{tr}[({\hat c}_{n}^{e,-})^{\dagger}{\hat
  c}_{n}^{e,-}]\rangle/N$ and similarly for the other terms. Also, the
averages $\langle \mbox{tr}(t'^{\dagger}t') \rangle=\langle
\mbox{tr}(r^{\dagger}r)\rangle=N/2$. From Eq. (\ref{arcoff}), the
matching equations at the left NIS-interface become
\begin{eqnarray}
f_{n}^{e,-}&=&\langle \mbox{tr}\left[ (b_n^e {\hat c}_{n}^{e,+}+a_n^h {\hat c}_{n}^{h,-}+c_n^e \delta_{n0}) \times c.c. \right] \rangle \nonumber \\
&=&B_n f_{n}^{e,-}+A_n f_{n}^{h,+}+C_nf_0\delta_{n0}, \nonumber \\
f_{n}^{h,+}&=&A_n f_{n}^{e,+}+B_n f_{n}^{h,-}+D_nf_0\delta_{n0},
\label{match2}
\end{eqnarray}
where the scattering probabilities\cite{Blonder82} $A_n=|a^{e/h}_n|^2$
and similarly $B_n$, $C_n$ and $D_n$, have been introduced. The total
distribution function for right (left) going electrons on the left
side at energy $E'$, $f^L_{\rightarrow(\leftarrow)}(E')$ is found by
summing up the distribution functions for right(left) going electrons
from all scattering states $\sigma$. From Eq. (\ref{match2}), it
follows that the total distribution functions at the left NIS-interface
are related as
\begin{eqnarray}
f^L_{\rightarrow}(E')&=&A(E')[1-f^L_{\leftarrow}(-E')]+B(E')f^L_{\leftarrow}(E') \nonumber \\
&+&T(E')f_0(E'), 
\label{match3}
\end{eqnarray}
where $T(E')=C(E')+D(E')=1-A(E')-B(E')$. In this derivation we also
used the symmetry for the total distribution functions
$f^e(E')=1-f^h(-E')$, which can be derived directly from Eqs.
(\ref{match1}) and (\ref{match2}). The relations between the total
distribution functions at the right interface and across the dot
follow in the same way. The resulting set of equations corresponds
exactly to the matching equations for distribution functions in Ref.
\onlinecite{Octavio83} (OTBK), with the important exception that the
scattering by the dot itself couples the left- and right moving
distributions of electrons in the normal part of the junction (i.e. a
SINI'NIS junction with transparency $1/2$ of $I'$). The current is
given by $\langle I \rangle=1/(eR_0)\int
dE'[f_{\rightarrow}(E')-f_{\leftarrow}(E')]$, with $R_0=h/(2e^2N)$.
This shows that the OTBK-approach can be rigorously justified for a
S-dot-S junction with broken time reversal symmetry in the dot.

We notice that the presented derivation is also applicable to a long
diffusive wire ($l \ll L$) SNS-junction with time reversal symmetry
broken in the wire. The only difference is that the middle barrier
(I') now has a transmission probability $l/L$.

In conclusion, we have studied the dc-current transport in a voltage
biased S-dot-S junction with an induced PE in the dot. It is found
that the PE is manifested as peaks in the conductance at voltages $eV
\approx 2(\Delta\pm E_{Th})/2n$.  These peaks are insensitive to
temperatures $kT \ll \Delta$ but are suppressed by a weak magnetic
field. The current for suppressed PE is independent of $E_{Th}$ and
magnetic field and is shown to be given by the OTBK-theory.

We thank J.  Kutchinsky for making the experimental data presented in
the paper available and we acknowledge helpful discussions with E.
Bezuglyi, J.  Kutchinsky, J. Bindslev-Hansen, J. Lantz and H.
Schomerus. This work was supported by TFR, NEDO and NUTEK.


\begin{thebibliography}{999}
\bibitem{excesscurr} The excess current is ranging from
$(G_Ne/\Delta)[16/3+\pi(9/2-4\sqrt{2})]$ for $\Gamma=1$ to $(G_Ne/\Delta)[\pi/4(7-4\sqrt{2})]$ for $\Gamma \ll 1$. In the limit $\Gamma \ll 1$ the distribution $\rho(D)$ is the same as for a dirty tunnel barrier and the result coincides with Ref. \onlinecite{Naveh00}.
\bibitem{numcom}
In all numerical curves shown, $M=20N$ and $1000$ Hamiltonians have been generated for each curve.
\end{thebibliography}
\end{document}